# Towards global equity in political polarization research

Max Falkenberg[1], Matteo Cinelli[2], Alessandro Galeazzi[3], Christopher A. Bail[4], Rosa M Benito[5], Axel Bruns[6], Anatoliy Gruzd[7], David Lazer[8], Jae K Lee[9], Jennifer McCoy[10], Kikuko Nagayoshi[11], David G Rand[12], Antonio Scala[2], Alexandra Siegel[13], Sander van der Linden[14], Onur Varol[15], Ingmar Weber[16], Magdalena Wojcieszak[17,18], Fabiana Zollo[19], Andrea Baronchelli[20,21,*], Walter Quattrociocchi[2,*]

*Corresponding authors: andrea.baronchelli.1@city.ac.uk , walter.quattrociocchi@uniroma1.it

## Abstract

With a folk understanding that political polarization refers to socio-political divisions within a society, many have proclaimed that we are more divided than ever. In this account, polarization has been blamed for populism, the erosion of social cohesion, the loss of trust in the institutions of democracy, legislative dysfunction, and the collective failure to address existential risks such as Covid-19 or climate change. However, at a global scale there is surprisingly little academic literature which conclusively supports these claims, with half of all studies being U.S.-focused. Here, we provide an overview of the global state of research on polarization, highlighting insights that are robust across countries, those unique to specific contexts, and key gaps in the literature. We argue that addressing these gaps is urgent, but has been hindered thus far by systemic and cultural barriers, such as regionally stratified restrictions on data access and misaligned research incentives. If continued cross-disciplinary inertia means that these disparities are left unaddressed, we see a substantial risk that countries will adopt policies to tackle polarization based on inappropriate evidence, risking flawed decision-making and the weakening of democratic institutions.

## Introduction

Researchers have gone to great lengths to understand the causes and consequences of political polarization in the United States, especially since Donald Trump's first election to the Presidency in 2016. However, a comparable evidence-base for understanding polarization outside the U.S. remains limited (Figure 1).

If insights from the U.S. were to generalize to other regions, the lack of localized evidence could, perhaps, be tolerated. However, as we will outline, there are numerous results from the U.S. which are outliers when viewed in a global context. Ensuring a globally relevant literature on political polarization is, or should be, a fundamental goal for the academic community (as several have argued previously [1], [2], [3]). Yet substantial systemic and cultural barriers exist. First, how polarization is studied is often influenced by norms established in U.S.-focused research (Box 1). Second, academic incentives are influenced by top interdisciplinary journals consistently favouring U.S.-focused research and reviews over work from other regions (SM). Finally, high quality polarization research is contingent on high quality data, but access to this data has become increasingly difficult, and is sometimes only available to U.S.-based researchers (SM).

[1] Central European University, Austria. [2] CNR-ISC, UoS Sapienza, Rome, Italy. [3] University of Padova, Italy. [4] Duke University, USA. [5] Universidad Politécnica de Madrid, Spain. [6] Queensland University of Technology, Australia. [7] Toronto Metropolitan University, Canada. [8] Northeastern, USA. [9] Sungkyunkwan University, South Korea. [10] Georgia State University, USA. [11] University of Tokyo, Japan. [12] MIT, USA. [13] University of Colorado Boulder, USA. [14] University of Cambridge, UK. [15] Sabanci University, Turkiye. [16] Saarland University, Germany. [17] University of Warsaw, Poland. [18] University of California San Diego, USA. [19] Ca' Foscari Venice, Italy. [20] City, University of London, UK. [21] The Alan Turing Institute, UK.

To address these issues, this article offers an overview of polarization research globally, highlighting key trends in ideological and affective polarization, the influence of socio-economic factors, and the impact of news and social media. Parts of this literature may seem disconnected, but this is a further reflection of the disciplinary fragmentation of polarization research.

Drawing from a growing set of examples, we discuss key results in the literature that are country-specific and those that, broadly speaking, generalize across a wide range of nations. We also highlight key areas where globally relevant literature is lacking, and provide recommendations for advancing towards a more globally equitable understanding of polarization. A failure to address these gaps is not only potentially dangerous, risking inefficient or counterproductive polarization interventions, but is intellectually complacent, further deepening existing inequities in our global understanding of polarization.

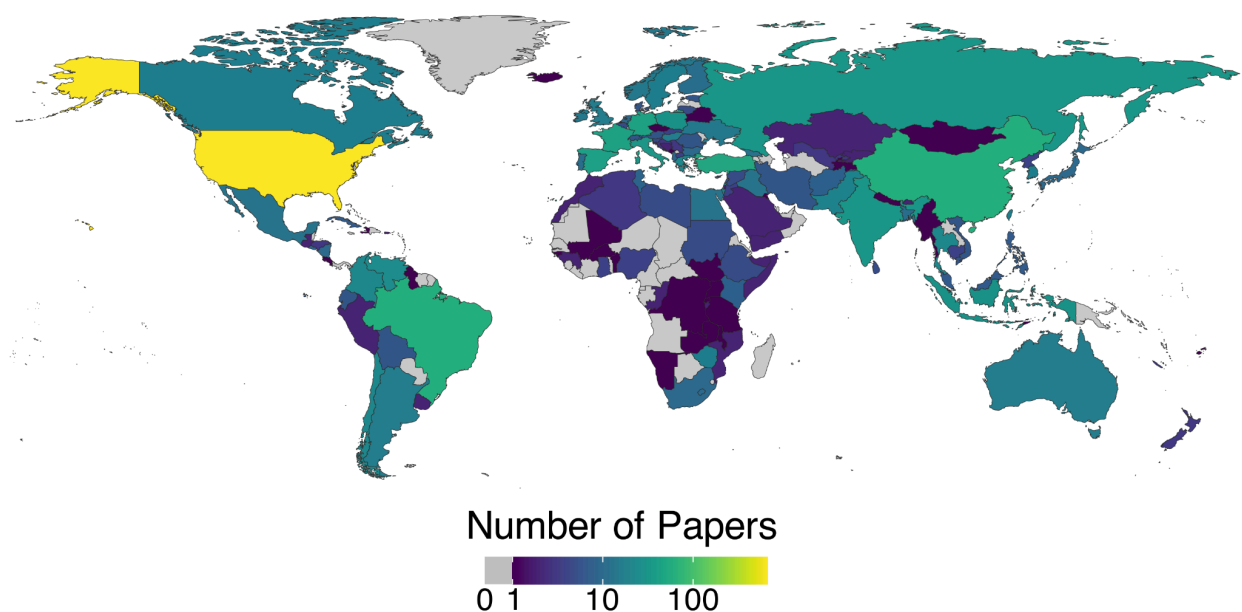

*Figure 1: The unequal distribution of polarization research.* Academic papers on political polarization which refer to a specific country overwhelmingly reference the United States. In total, 2,341 papers use the term "political polarization" (or "political polarisation") in their title or abstract of which 1,309 refer to at least 1 country. The United States is mentioned in 650 papers (49.7% of papers with a country mention). Many papers which do not mention a country are U.S. focused, but do not state so explicitly. See SM for details.

**Box 1: The normative U.S. framing of polarization.** In selecting "polarization" as their word of 2024, Merriam-Webster defined polarization as "division into two sharply distinct opposites;

especially, a state in which the opinions, beliefs, or interests of a group … become concentrated at opposing extremes". While such a framing may be appropriate for the public's understanding of the U.S.'s unusual Manichean two-party democracy, there are many cases where such a framing may not be appropriate. For instance, it is far from clear how polarization should be understood and measured in multi-party democracies (recent work has started to address this issue [3]; also SM), or in the absence of pluralist party systems where the opposition to an authoritarian regime is hard to measure (SM). Similarly, there is substantial work discussing how polarization can be formulated without explicit reference to the extremes, changing how the concept is interpreted [4]. While such issues can seem abstract, they can directly affect the global context for understanding polarization when, for example, cross-country political surveys ask individuals to place themselves ideologically on a one-dimensional left-right scale, when other axes of political belief may be equally important (e.g., the establishment-anti establishment dimension; SM).

## Global trends in polarization

Historically, scholars have focused largely on the study of ideological polarization, typically understood as the gap between liberals and conservatives, and by extension between the political left and right (SM). In contrast, since the late 2000s, evidence from the U.S. that animosity between political opponents has grown disproportionately relative to policy differences has led researchers to shift focus away from ideological polarization and towards affective polarization ([5], SM): the phenomenon whereby an individual has negative feelings towards their political out-group (and/or positive feelings towards their in-group).

Amongst political elites, comparative data from 21 high-income countries reveals a small but significant increase in elite ideological polarization, with Canada and Australia cited as countries with low ideological polarization, in contrast to Spain, Italy and Sweden where ideological polarization is high (SM). However, at a global scale, ideological polarization does not appear to have increased amongst the mass public ([6], SM). This is in contrast to the trend observed in the United States where growing ideological polarization (both mass and elite) is now well documented (SM). Such increases are, however, typically smaller than the variability observed across countries ([6], SM).

In the United States, the state of affective polarization has been described as "political sectarianism" [5], drawing parallels to religious and ethnic sectarianism, and having often been linked to the rise of populism (SM). Referencing data from 12 OECD countries, Finkel et al. argue that the state of U.S. affective polarization is exceptional, with the U.S. having suffered the largest increase in affective polarization over the last 40 years [5], [7]. However, estimates of the absolute levels of affective polarization from 53 countries point towards countries such as Türkiye, Hungary, and Kenya as examples of high affective polarization, and the Philippines, the Netherlands, Belgium, and Japan as examples of countries with low levels of affective polarization, with the U.S. falling in the middle of the pack [8]. For many countries (including authoritarian states, where measuring polarization is particularly challenging; SM) quantitative estimates of affective polarization are still unavailable (SM).

# The relationship between ideological and affective polarization

In the United States, the relationship between ideological and affective polarization is debated, rooted in the question of whether policy differences drive partisan animosity (SM). This debate draws parallels globally, where the evidence for their relationship is mixed. Ideological and affective polarization are broadly correlated [9], but some countries display significant ideological polarization without high out-party animosity (e.g., the Netherlands), while others are afflicted with severe affective polarization despite the principal parties being largely moderate (e.g., Montenegro and Serbia). Importantly, survey evidence from 53 countries suggests that high levels of affective polarization are correlated with democratic backsliding, but high levels of ideological polarization are not ([8], SM).

The reasons for country-level differences in the relationship between ideological and affective polarization are complex: The U.S.-centric literature has made a two-party "us-vs-them" framing of polarization the norm (Box 1), but in multi-party states it is not uncommon for individuals to have positive (or negative) feelings towards multiple parties (e.g., in Finland or Sweden [3]). Not only does this complicate how to measure polarization, it also introduces new political dynamics which can have both positive and negative impacts, for instance, related to the sustainability and effectiveness of coalition governments [9].

In this context, research suggests that affective polarization is higher in majoritarian political systems [10] where power is concentrated in the hands of a single party (e.g., France, UK) than in systems based on consensual governance (e.g., Switzerland, the Netherlands), while the opposite trend is observed for ideological polarization (SM).

Socio-economic factors (SM) and income inequality are also implicated as key drivers of polarization. Evidence from 20 Western democracies (Europe, USA, Canada, New Zealand, Australia) shows that affective polarization is worse when levels of income inequality and unemployment are high [10]. But, again, countries with high income inequality show lower levels of ideological polarization amongst the mass public (e.g., the U.S.) than lower income inequality countries (e.g., Sweden; SM).

Finally, it is often assumed that polarization has roots in ethnic or religious differences. However, while these factors may correlate with observed differences in polarization, and may be used as the basis of polarizing strategies by political leaders (e.g., in the U.S., Hungary, or Türkiye), multiple comparative studies have shown that social cleavages do not, inherently, predict damaging forms of polarization (SM).

# Polarization and information exposure

## The role of the news media.

Partisan news media are often blamed for increasing polarization, but evidence to support this assertion is mixed and beset by technical issues and geographic limitations. Again, the available literature is heavily U.S.-centric [11] and the role of the news media in developing countries or in autocracies is rarely considered.

Most studies on the polarizing role of the news media concern either the supply side of the news – how the news media ecosystem is aligned with specific political ideologies – or the reception side – namely what news individuals consume. Addressing the former, Western news media systems have been categorized according to "political parallelism", referring to the one-to-one correspondence between news outlets and political parties (SM). Using this framework, countries were classified as having low levels of political parallelism (e.g., the U.S., UK, Ireland), high levels but lowering over time (e.g. Austria, Belgium, Finland, Germany), or sustained high levels (e.g., Italy, France, Spain, Greece), revealing substantial differences in the politicization of the news media across developed countries.

With regard to the consumption side, the proliferation of the internet has enabled easy access to increasingly diverse online content, some of which is highly partisan or radical (SM). Nevertheless, survey self-reports in 12 developed countries [12] suggest that the majority consume largely centrist and moderate news, both on- and offline, despite (largely unfounded) fears that social media algorithms and "filter bubbles" will drive users to more radical content (SM). The U.S. has the most partisan consumption, and individual communities do have a highly skewed news media diet (more so in the U.S. than other countries; SM), but absolute levels of partisan news consumption remain low. It is also not clear how to compare these results to countries where the politicization of the news media is more explicit, including authoritarian states where news outlets are forced to support the regime in power or risk political backlash (SM).

Questions of news supply and demand do not answer whether the news media impact polarization in the general population. Limited survey-based and experimental evidence from the U.S. and Israel suggests that exposure to partisan news outlets and their messages reinforces prior political attitudes and affective polarization (SM). However, given the challenges in measuring exposure to (partisan) media using self-reports and in ascertaining their effects outside experimental settings, this research may not portray the role of the news media in the "real world" accurately (SM). Additionally, the global loss of trust in the news, and very low engagement with news media content in general, mean that researchers are increasingly questioning the polarizing role of the news media (SM).

## The role of the internet & social media.

The internet and social media continue to play a dominant role in how citizens consume content. Despite this, claims of social media's negative impacts are often anecdotal, or based on correlational evidence [13], and there are substantial differences across countries in the extent to which social media is seen as democratically harmful (Americans see social media as far more damaging than other surveyed populations; SM).

With a focus on user-user interactions, one perspective argues that social media drive polarization by enabling politically engaged users to primarily interact with other ideologically aligned individuals – a pattern observed across many platforms and countries – indicative of an epistemic bubble, or echo chamber, where ideological views and out-group affect may become more extreme (SM).

However, another perspective argues that these homophilous interaction patterns do not constitute ideological isolation. First, social media interactions in most countries are not primarily with political actors or content (SM). Second, when individuals interact with news content rather than individuals, their news diet is often politically moderate (see above). Third, while certain interaction types (like reposts) give the appearance of isolated communities, other interaction types (mentions or follows) reveal that out-group interaction is not so rare (SM), albeit these interactions are often toxic attacks (a feature common across many countries; SM). Finally, even if online interactions are ideologically homogeneous, individuals' may in fact be more isolated in their offline environments (SM), a factor likely to be highly country- and context-dependent.

Whether biased information exposure on social media leads groups to adopt more extreme views is debated (SM). The largest Facebook field experiment to date – a collaboration with Meta which was limited to U.S. researchers (SM) – finds that reducing users' exposure to content from like-minded sources does not measurably affect affective polarization or ideological extremity [14]. However, similar research (with smaller study sizes) suggests that such results are likely to be country-dependent: One U.S. study found that deactivating Facebook reduced both ideological and affective polarization, while a replication in Bosnia-Herzegovina found an increase in polarization (SM), with evidence to suggest that this unexpected increase may be related to individuals' lack of offline exposure to other ethnicities. Critically, the question of whether social media is polarizing is linked to questions regarding the role of social media algorithms, the growing influence of AI, and how the internet is regulated. While our understanding of these factors is still at a very early stage, it is clear that they are likely to impact the polarizing role of the internet in different ways across countries (SM).

# Moving forward: The global challenge of countering polarization.

Preventing the potential damaging societal impacts of political polarization requires effective and long lasting interventions. However, the current dominance of U.S. research has put an outsized

*Falkenberg et al., Towards Global Equity in Political Polarization Research, (2025)*                         6

emphasis on developing interventions for affective polarization between Democrats and Republicans in the U.S. [15] (SM), with few interventions having been tested elsewhere. Relying on such narrowly designed measures risks introducing inappropriate or even counterproductive solutions in diverse societal and cultural contexts.

Fortunately, scholars are increasingly aware of the need to study interventions globally, with several projects already underway (for example [globalsocialmediastudy.com](globalsocialmediastudy.com) and [templetonworldcharity.org/LLPW](templetonworldcharity.org/LLPW)). However, the impact of these studies risks remaining limited without systemic and cultural changes in how polarization research is produced and communicated.

Researchers working on understudied regions often face career disadvantages due to the perceived lack of visibility, relevance, and therefore impact, of their work. Academic institutions, publishers, and policymakers must actively promote and reward work that brings global balance to polarization research. This is not a call for less research on the U.S., which will remain well-funded and prolific, but rather to increase recognition and support for research that expands our understanding to include diverse global perspectives.

A first step to improving the current state, given the constraints on academic resources, is to prioritize comparative research that identifies both shared and unique aspects of polarization across different countries. As part of this, it is crucial that country-specific research avoids over-generalizing their results. For example, journals could mandate that studies focusing solely on the U.S. explicitly state this in their title or abstract (many currently don't), as is commonplace for research focusing on almost any other country.

To prevent further global disparities in polarization research, we argue that researchers must push back against growing restrictions on data access now. Since 2023, several social media companies (including Meta, X, Reddit) have shuttered academic access to their platforms' data, and new opportunities for data access such as the European Union's Data Services Act remain largely unproven (SM). Overcoming these challenges is essential for adopting truly global approaches to studying polarization.

Finally, as our understanding of polarization evolves and new interventions are developed, both academics and policymakers must account for the unique regulatory and practical challenges of each country (SM). Interventions must be both ethically and practically tailored to the local context. We cannot assume that societies outside the U.S. – where empirical evidence is comparably limited – face the same problems which can be solved with the same solutions. Academics play a critical role in gathering evidence to better understand how polarization manifests across countries, but the implementation of interventions is necessarily political and often controversial. These decisions must rest with local stakeholders and decision makers.

# Supplementary Material

***Note that reference numbers below refer to the supplementary references (end of this document), not to the references listed in the main text above.***

1. **"Second, academic incentives are influenced by top interdisciplinary journals consistently favouring U.S.-focused research and reviews over work from other regions."**

   Examples of prominent reviews which focus on the U.S. only: [1], [2], [3], [4]. Examples of prominent research articles recently published in top interdisciplinary journals which focus on the U.S. only: [5], [6], [7].

2. **"Finally, high quality polarization research is contingent on high quality data, but access to this data has become increasingly difficult, and is sometimes only available to U.S.-based researchers."**

   See recent discussions on the need for data availability, and why new measures to improve data availability, such as the European Union's Digital Services Act, might not succeed due to poor compliance by tech companies: [8], [9], [10].

   Note that the Meta election studies [6], [7], [11], [12] looking, amongst other things, at the polarizing impact of Meta's platforms was restricted to U.S.-based researchers. For a discussion of the collaboration, see [13].

3. **"Figure 1: The unequal distribution of polarization research."**

   Data to produce Figure 1 taken from papers indexed by Semantic Scholar. Includes all papers written in English which use the term "political polarization" or "political polarisation" (American and British English spellings) in their title or abstract, indexed up until December 2023. In total, 2,341 papers use the term "political polarization" in their title or abstract of which 1,309 refer to at least 1 country. The United States is mentioned in 650 papers (49.7% of papers with a country mention). China and Brazil, the second and third most referenced countries, are mentioned in 60 (4.6%) and 59 (4.5%) papers/abstracts respectively. Contributions to the Chinese total include papers on Covid-19, Taiwan, and Hong Kong.

4. **"For instance, it is far from clear how polarization should be understood and measured in multi-party democracies."**

The U.S. is the archetypal example of a two-party democracy, with this two-party structure informing how polarization is measured. But for states with multiple political parties this two-party framing of polarization may not be a natural fit. Researchers have adapted conventional two-party measures of polarization for use in multi-party contexts [14], [15], but, to some extent, it remains an open question as to whether key concepts in the study of polarization change when extended from a one-dimensional political spectrum to a multi-dimensional context.

For example, the standard adaptations of ideological and affective polarization to countries with three or more parties continues to treat polarization as a one-dimensional us-vs-them phenomenon since each pair of parties is handled independently, rather than treating polarization as an inherently multipolar concept. It is also not clear whether such an approach could be used in authoritarian countries without a measurable opposition. Approaches which try to retain polarization's multipolar nature are being developed, but remain largely confined to studies in network science and social media (for example [16], [17]).

5. **"For instance, it is far from clear how polarization should be understood … in the absence of pluralist party systems where the opposition to an authoritarian regime is hard to measure."**

Despite its normative framing as a democratic phenomenon, polarization is an important concept for authoritarian states [18], [19], [20]. As in democracies, individuals in authoritarian societies have diverse ideological views and may have positive or negative affect towards various groups (e.g., towards supporters of the regime in power). But studying polarization in these contexts involves unique challenges (see below). Beyond its inherent value, the study of polarization in these states can offer valuable insights for new democracies which retain a legacy of authoritarianism (e.g., Kenya [21] and Liberia [22]) and can guide scholars interested in polarization's role in eroding democracy [23], [24], [25], [26], [27].

*Challenges*: Particularly in authoritarian states, polarization can be exploited as a tool to cement power and discredit opponents [28]. Researchers can play an important role in understanding political dynamics in these contexts. For example, research has shown how the increased polarization of online interaction patterns preceded periods of violence during the Arab Spring [29], how the Chinese government produces up to half a billion social media posts each year to distract citizens from political topics [30], and how politicians in Zimbabwe deliberately exploited polarization to reshape boundaries of conflict and mobilize constituencies [19].

Crucially, given the risks posed to research subjects in authoritarian states – including severe physical repression [20], or exile [31] – researchers must address a number of practical and ethical challenges to studying polarization:

- Respondents in authoritarian contexts are "hard to survey" [32], meaning that studies suffer from numerous biases including the censorship of data (e.g., in

- China [33]), preference falsification in surveys, and coordinated inauthentic activity. These biases may be mitigated by using list experiments or conjoint experiments [34]. Additionally, preference falsification may be counteracted using "anonymity sampling" [35] which improves survey accuracy while protecting respondent anonymity.
- Scientists are increasingly emphasizing the need for publicly available data for study replication. However, in authoritarian contexts, such data may put survey respondents at risk if not carefully anonymised. Researchers must take extra care when providing public replication datasets – including the ids of social media posts – that contain politically sensitive content. The use of secure data archives may offer a path for data sharing without putting research subjects at risk [36].
- Even if data are anonymised, scholars should be conscious that their work may be used as "opposition research", and indirectly benefit the regime in power. For example, papers with real-time monitoring of opposition groups could be used as justification for cracking down on protesters. An example which highlighted these issues is the ENCORE study, in which internet censorship was measured at the user level across a number of countries – including China, Iran, Pakistan, and Saudi Arabia – but without individual consent [37].

6. **"While such issues can seem abstract, they can directly affect the measurement of polarization when, for example, cross-country political surveys ask individuals to place themselves ideologically on a one-dimensional left-right scale, when other axes of political belief may be equally important (e.g., the establishment-anti establishment dimension)."**

Important survey studies carried out across a large number of countries which ask a classic left-right ideology question include the Comparative Study of Electoral Systems (https://cses.org/) and the World Values Survey (https://www.worldvaluessurvey.org/wvs.jsp).

For a discussion of dimensions of political belief which may be as important as the classic left-right ideological dimension, see Refs. [38], [39], [40].

7. **"Historically, scholars have focused largely on the study of ideological polarization, typically understood as the gap between liberals and conservatives, and by extension between the political left and right."**

See Refs. [41], [42], [43].

8. **"In contrast, since the late 2000s, evidence from the U.S. that animosity between political opponents has grown disproportionately relative to policy differences has led researchers to shift focus away from ideological polarization and towards affective polarization."**

See Refs. [3], [44], [45].

9. **"Amongst political elites, comparative data from 21 high-income countries reveals a small but significant increase in elite ideological polarization, with Canada and Australia cited as countries with low ideological polarization, in contrast to Spain, Italy and Sweden where ideological polarization is particularly high."**

   See Ref. [46].

10. **"At a global scale, ideological polarization does not appear to have increased amongst the mass public."**

    See Ref. [47] where ideological polarization is computed as an aggregate of various issue positions across 105 nations. Similar conclusions are reached by computing the variance (a common measure used for extracting a polarization score from a distribution of opinions; see Refs. [46], [48]) of the self-reported left-right ideological scores across countries using CSES data from 59 countries [49].

11. **"This is in contrast to the trend observed in the United States where growing ideological polarization (both mass and elite) is now well documented."**

    See Refs. [50], [51].

    It is also worth noting that the measured level of affective polarization depends on whether the focus is on members of opposing political parties in general, or political elites and leaders from opposed parties. While most studies of affective polarization consider how partisans feel about members of their out-party, Reiljan et al. [52] show that the U.S. is the only country from 40 democracies where affective polarization towards party leaders significantly exceeds affective polarization towards members of the out-party.

12. **"Such increases are, however, typically smaller than the variability observed across countries".**

    See Refs. [3], [44], [45].

13. **"In the United States, the state of affective polarization has been described as "political sectarianism", drawing parallels to religious and ethnic sectarianism, and having often been linked to the rise of populism."**

    The global rise in populism [53] is often seen as a symptom of rising political polarization. However, as Norris [54] highlights, populist parties across the world show significant diversity in their ideological positions, suggesting that populism may be better characterized by shared political rhetoric and strategies, rather than by a common ideology. Despite this, there is evidence from Europe to suggest that populist parties are

associated with higher levels of affective polarization, particularly on the radical right [55].

Some scholars have linked the rise of populism to the loss of trust in political institutions [56], while others have argued that populism has deeper roots in the emergence of anti-establishment attitudes and individuals' "need for chaos" [38], [39]. The latter is particularly important because it reframes political cleavages as existing between the political elite and a politically disengaged, anti-establishment public, rather than between the political left and right. The importance of this anti-establishment dimension is clearest in the U.S. following Trump's election to the U.S. Presidency, but political movements in other countries have also seized on this sentiment, notably in states with high affective polarization like Türkiye and Hungary.

14. **"For many countries (including authoritarian states, where measuring polarization is particularly challenging)..."**

One concept which is used to assess damaging forms of polarization, including in authoritarian states and states with hard to survey populations, is pernicious polarization – a metric which has received relatively little interdisciplinary attention – where country-experts, not the local population, are asked to assess whether the country in question has divided into mutually antagonistic camps [57]. This measure points towards Brazil, Mexico, the United States and India as examples of democracies with particularly high levels of pernicious polarization and reveals that every world region other than Oceania has suffered an increase in pernicious polarization since 2005, albeit levels were much higher in Europe and the Americas during the first half of the twentieth century. However, pernicious polarization is only one of many polarization measures. The academic community has not yet reached a consensus as to how different polarization types are related, the contexts in which it is better to study one polarization type over another, and whether polarization is even undesirable [58]. This complicates how to interpret the phenomenon. For instance, it is not obvious how to reconcile the high level of pernicious polarization in the U.S. with the relatively moderate levels of affective polarization.

15. **"For many countries… quantitative estimates of affective polarization are still unavailable."**

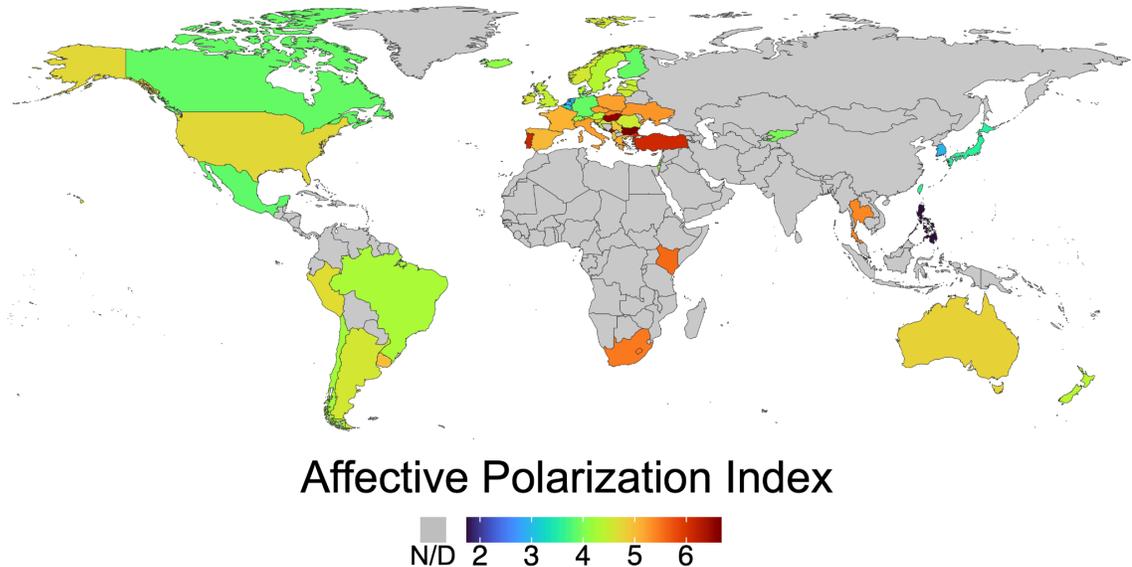

*The affective polarization index (API) which measures the emotional gap that partisans feel towards their in-party, as opposed to their out party. For many countries in Africa and Asia, an estimate of affective polarization is not available. Data from Ref. [25]. Note that the calculation of the API uses data from the Comparative Study of Electoral Systems (CSES) on political affect which requires the identification of at least two political parties in a country, and weights partisan groups according to a party's vote share. As such the measure is not easy to adapt to authoritarian states which do not have multiple political parties.*

16. **"In the United States, the relationship between ideological and affective polarization is debated, rooted in the question of whether policy differences drive partisan animosity."**

In the U.S., some scholars have argued that ideological and affective polarization are largely distinct [44], [45], while others have identified ideological polarization as a key driver of partisan animosity [59], [60]. Proponents of the former view often link affective polarization to partisan sorting: the growing alignment between individuals' party political affiliation (Republican or Democrat) and ideological identity (conservative or liberal; see ref. [3]), as well as to various social identities including race, ethnicity, religion, class, socioeconomic status, and location [45], [61], [62]. Importantly, while proponents of this view identify political identity as a critical factor in driving partisan animosity, they, in general, are not claiming that policy views do not continue to play an important role in explaining affective polarization.

The partisan sorting hypothesis is important because it frames partisanship in terms of identity rather than on the basis of policy positions – a key change in the conceptual framing of polarization – a feature of politics also observed outside the U.S., including in the UK, Sweden, and the Netherlands [63]. However, recent research suggests that within-party affective polarization (i.e., negative affect aimed at other groups in one's own party) can be as strong as between-party affective polarization, thereby implying that political faction may be more important than party identity in driving out-group affect [64].

17. **"Importantly, survey evidence from 53 countries suggests that high levels of affective polarization are correlated with democratic backsliding, but high levels of ideological polarization are not."**

    It is debated whether ideological polarization should be viewed as democratically problematic or not. In the 1950s, political scientists suggested that the United States was not ideologically polarized enough, arguing that a failure by the Republican and Democratic parties to offer meaningful differences in policy could risk voters turning away from democracy and towards more radical political movements [65]. However, excessive ideological polarization – often defined heuristically based on the associated negative outcomes – is seen as equally damaging, preventing political compromise, generating political deadlock (e.g., following the Brexit vote in the UK), and facilitating the rise to power of extreme political factions.

    More generally, there remains a lack of consensus as to what constitutes good or bad polarization. This obfuscates what interventions should counter exactly, with some even questioning whether affective polarization should really be seen as a negative [58]. Furthermore, given some evidence that affective polarization is not a strong predictor of anti-democratic attitudes [66], [67], it is possible that concentrating efforts on mitigating polarization may not translate into other outcomes that are key to more effective democracy.

18. **"In this context, research suggests that affective polarization is higher in majoritarian political systems where power is concentrated in the hands of a single party (e.g., France, UK) than in systems based on consensual governance (e.g., Switzerland, the Netherlands), while the opposite trend is observed for ideological polarization."**

    As highlighted by Ref. [46], one of the best examples of how the political system affects ideological polarization is the example of New Zealand: When the electoral system changed to proportional representation in 1993, the measured level of ideological polarization doubled [68].

19. **"Socio-economic factors, and income inequality in particular, are also often implicated as a key driver of polarization."**

    For discussions on socio-economic factors driving polarization see Refs. [46], [69], [70], [71], [72], [73], [74], [75].

    Several other factors may alter how polarization manifests, but it is often hard to establish causal evidence of a relationship. Important examples include differences in

political engagement across countries [69], the role of supra-national organizations (e.g., the European Union, or Nato [76]), and differences in news media and internet use.

20. **"But, again, countries with high income inequality show lower levels of ideological polarization amongst the mass public (e.g., the U.S.) than lower income inequality countries (e.g., Sweden)."**

    See Refs. [70], [75].

21. **"Finally, It is often assumed that polarization has roots in ethnic or religious differences. However, while these factors may correlate with observed differences in polarization, and may be used as the basis of polarizing strategies by political leaders (e.g., in the U.S., Hungary, or Türkiye), multiple comparative studies have shown that social cleavages do not, inherently, predict damaging forms of polarization."**

    It is often assumed that polarization has roots in ethnic or religious differences. However, while these factors may correlate with observed differences in polarization [47], multiple comparative studies have shown that social cleavages do not, inherently, predict damaging forms of polarization [27], [57], [77], [78].

    Ethnic, religious or social class divisions become politically relevant cleavages when people develop shared identities and values, and organizations like parties or unions represent them [79], [80]. And these cleavages become polarizing when political actors make them the basis of polarizing strategies, for instance, through the mobilization of anti-immigration sentiment in Europe and the U.S. [81]. But as Benson [82] points out, while the social psychology of intergroup conflict may explain the intensification of the in-group identification and out-group antipathy of affective polarization, it does not explain how those identities become politically salient in the first place. For this, scholars turn to elite agency and the divisive rhetoric of polarizing politics that politicizes certain social cleavages, grievances and political identities [19], [83].

    For example, ethnic and religious divides were politicized in India and Türkiye as the secular solution of the founders gave way to Hindu nationalism under Modi's BJP in India and the Islamist movement and Kurdish rift under Erdogan's AKP in Türkiye [84], [85]. On the other hand, some countries with significant ethnic divides have not descended into an ethnicity-defined political rift [86]. A comparative study of four Asian countries finds that those with clearly identifiable social cleavages, Taiwan and Indonesia, suffered only short episodes of damaging polarization, while those with less identifiable social cleavages, Thailand and the Philippines, suffered more destructive consequences, including democratic collapse in the Thai case [87].

    Finally, certain cleavages may be more prone to damaging polarization, such as the unresolved, emotionally-charged historic debates around citizenship, national identities, and founding myths that Somer and McCoy [88] call "formative rifts". Examples include

the claim that political legitimacy is to be conferred only on those with national liberation war experience in Zimbabwe [89], or the legacy of unequal citizenship rights and political exclusion around race, gender and religion in the U.S. [90], [91].

22. **"Addressing the former, Western news media systems have been categorized according to "political parallelism", referring to the one-to-one correspondence between news outlets and political parties."**

    See Ref. [92] and updated versions of the classification with largely comparable results in Ref. [93].

23. **"With regard to the consumption side, the proliferation of the internet has enabled easy access to increasingly diverse online content, some of which is highly partisan, or radical."**

    See Refs. [94], [95], [96].

24. **"Despite (largely unfounded) fears that social media algorithms and "filter bubbles" will drive users to more radical content."**

    Some have argued that the internet plays a key role in supplying partisan news by selectively filtering the sources which individuals are exposed to (sometimes referred to as filter bubbles [97]). However, most studies do not find evidence to support the filter bubble hypothesis [98], [99]. For instance, data from the U.S. [100] and Spain [101] show that Google search has a moderating effect on the consumption of partisan news, relative to individuals' own tendency to self-select news sources.

25. **"Individual communities do have a highly skewed news media diet (more so in the U.S. than other countries)."**

    See Ref. [102]. See also a recent study that shows that there are a group of U.S. Facebook users who consume exclusively right-leaning news [7].

26. **"It is also not clear how to compare these results to countries where the politicization of the news media is more explicit, including authoritarian states where news outlets are forced to support the regime in power or risk political backlash."**

    See Ref. [103].

27. **"Limited survey-based and experimental evidence from the U.S. and Israel suggests that exposure to partisan news outlets and their messages reinforces prior political attitudes and affective polarization."**

See Refs. [96], [104], [105], [106].

28. **"However, given the challenges in measuring exposure to (partisan) media using self-reports and in ascertaining their effects outside experimental settings, this research may not portray the role of the news media in the "real world" accurately."**

A major challenge in assessing the polarizing role of the news media is acquiring high quality causal evidence [107], especially given the difficulty of establishing a baseline for the content survey respondents are exposed to in their day-to-day lives. Partly addressing this challenge, a 2020 study incentivised regular Fox News viewers to watch CNN for a month, finding that this switch had a moderating effect on their political attitudes [108], thereby implying that continued exposure to Fox News does have some polarizing effects. Yet, other work finds null effects from exposure to politically aligned news media: Online behavioral data shows that partisan news consumption has little impact on attitudes or affective polarization in the U.S. [11], [109], [110], but how these results translate to population level polarization, and to other world regions, is unclear.

29. **"Additionally, the global loss of trust in the news, and very low engagement with news media content in general, mean that researchers are increasingly questioning the polarizing role of the news media."**

Recently, some have questioned the polarizing role of the news media, given that few people are interested in or regularly consume the news. In fact, self-surveys carried out across 46 countries (in Europe, the Americas, Asia, and some in Africa) show that interest in the news has fallen significantly in the last decade [111]. No country has seen a growth in interest, and few countries have maintained stable levels – notably Finland, the Netherlands, and South Korea. Providing direct evidence of this, studies using web browsing data show that only 3.4% of all URLs visited by large samples in the U.S., Poland, and the Netherlands are news, and, even when individuals visit partisan domains, more than half of the content they consume is apolitical, such as sports or food recipes [109]; see also [112]. On social media – which has become a leading gateway for accessing news content [111] – news makes up only a small part of a user's feed (less than 3% for U.S. Facebook users). This lack of interest in the news and political content may partly explain diverging trends in mass and elite polarization across countries (see pages 4-6).

Rather than reducing exposure to political content, some argue that increasing public exposure to high quality news may have a depolarizing effect. This would pull more moderate citizens into the democratic process, increase public resilience to mis- and disinformation [113], [114], [115], and decrease their susceptibility to populist or hyper-partisan rhetoric [95]. However, the widespread decline in trust in the news (16 out of 19 surveyed countries [111]) casts a shadow on this approach. Strong public service broadcasters may play a key role in maintaining trust, but, in many markets, politicians and other public figures are actively undermining trust through their continued criticism of the news media, e.g., in Greece, Peru, the U.S., and the UK [111].

30. **"Americans see social media as far more damaging than other surveyed populations."**

   See Ref. [116].

31. **"With a focus on user-user interactions, one perspective argues that social media drive polarization by enabling politically engaged users to primarily interact with other ideologically aligned individuals – a pattern observed across many platforms and countries – indicative of an epistemic bubble, or echo chamber, where ideological views and out-group affect may become more extreme."**

   Online, polarization is most often framed through the lens of individuals' network-based interactions (sometimes referred to as interactional polarization [117], [118]). Here, political ideology is inferred in reference to specific political narratives or identities, known political actors, or partisan media outlets [16], [17], [119], [120], [121].

   These homophilous user-user interaction patterns are found near-universally across countries including in the U.S. [119], [122], Brazil [123], Egypt [29], Israel [117], Venezuela [124], Pakistan [125], several Western countries [118], and Japan [126]. This is also true across a number of mainstream platforms [127], [128], but to varying extents. For example, partisan homophily is less pronounced on Reddit than on Twitter or Facebook [127], [129].

   For a discussion of echo chambers and epistemic bubbles see Ref. [127], [130], [131], and for how they may make individuals' views more extreme see Ref. [132].

32. **"First, social media interactions in most countries are not primarily with political actors or content."**

   See Refs. [122], [133].

33. **"Second, when individuals interact with news content rather than individuals, their news diet is often politically moderate."**

   See Refs. [102], [110], [134].

34. **"Third, while certain interaction types (like reposts) give the appearance of isolated communities, other interaction types (mentions or follows) reveal that out-group interaction is not so rare, albeit these interactions are often toxic attacks (a feature common across many countries)."**

See Refs. [118], [131], [135], [136].

35. **"Finally, even if online interactions are ideologically homogeneous, individuals' may in fact be more isolated in their offline environments."**

See Refs. [71], [137], [138]. Note that for these reasons, some scholars have suggested that research should look beyond the mere act of interaction and instead focus on interaction quality as well as on promoting prosocial behavior [139], [140], [141], [142].

36. **"Whether biased information exposure on social media leads groups to adopt more extreme views is debated."**

Assessing whether social media drives individuals to more extreme views is debated in part due to the difficulty in acquiring high quality causal evidence of the individual-level effects of platforms and content. A study in the U.S. using fictional social media profiles showed that exposure to opposing views reduced participants' polarization on specific issues [143]. However, a Twitter field experiment which exposed U.S. partisans to out-party messages resulted in an increase, rather than decrease, in ideological polarization [144]. In the largest field experiment on Facebook completed during the 2020 U.S. elections, Nyhan et al. [6] found that reducing users' exposure to content from like-minded sources did not measurably affect affective polarization or ideological extremity.

While the presence of isolated information environments on mainstream social media is debated, recent years have seen the emergence of alternative online spaces catering to specific political ideologies – e.g., Parler, Gab, Voat, or Gettr. Here ideological isolation is more likely, possibly reinforcing any radical views already held by their users, but evidence to support this concern is mixed [145], [146]. Perhaps more importantly, the polarizing effect of these platforms may be minimal at the population level given evidence that alternative social media struggle to retain active users when politically diverse mainstream platforms are available instead [146]. Outside the U.S., a number of social media platforms have established themselves catering to specific national populations rather than ideologies (e.g., Sina Weibo in China and VKontakte in Russia), but their role in polarizing their users remains poorly understood, and they are subject to substantial state manipulation complicating their analysis.

37. **"A collaboration with Meta which was limited to U.S. researchers."**

See discussion in Ref. [13].

38. **"However, similar research (with smaller study sizes) suggests that such results are likely to be country-dependent: A U.S. study found that deactivating Facebook** **reduced both ideological and affective polarization, while a replication in Bosnia-Herzegovina found an increase in polarization (SM), with evidence to suggest that this unexpected increase may be related to individuals' lack of offline exposure to other ethnicities."**

See Ref. [147] for the U.S. study, and Ref. [148] for the Bosnian study. Note that the Bosnian study did not replicate in another ethnically polarized setting, Cyprus [149].

39. **"Critically, the question of whether social media is polarizing is linked to questions regarding the role of social media algorithms, the growing influence of AI on the internet, and how the internet is regulated. While our understanding of these factors is still at a very early stage, it is clear that they are likely to impact the polarizing role of the internet in different ways across countries.**

    There are broad concerns that the polarizing effect of social media may be exacerbated by recommendation algorithms, AI generated content, and bots [95], [150], [151], [152].

    Since the quantity of online content vastly exceeds what any individual can consume, social media companies must – by necessity – curate the content which their users are exposed to. For example, on Youtube recommended content makes up over 70% of the videos watched [153]. However, with AI-based recommendation systems optimizing for user engagement [154], there are fears of a self-reinforcing cycle in which consumers are exposed to increasingly homogeneous content which – if the content is political in nature – may drive polarization.

    Unfortunately, the commercial interests of tech companies mean that independently auditing this polarizing role is difficult. From the few studies available, U.S. data on Instagram and Facebook suggest that various changes to their recommendation systems do not impact ideological or affective polarization. These changes included removing re-shared content [11], removing all content rankings [12], and down-ranking content from ideologically aligned sources [6]. On Youtube, data suggests that recommendations are largely reflective of individual preferences, and that the algorithm only has a mild role in driving users to more politically moderate content [155]. However, data from Twitter shows that their recommendation system amplified content from the political right more so than from the political left in six of seven high-income countries [156].

    We may expect that social media recommendation systems are largely country independent. However, their inputs are context specific and therefore country specific. These include users' IP addresses – revealing location – the languages they use, and the regional topics they interact with. Unfortunately, there is currently insufficient research to fairly assess how these regional factors will impact the polarizing role of social media algorithms, but one recent study does show that, across four culturally

distinct countries, public understanding of how social media algorithms operate varied substantially by country and demographic [157].

Similar fears are raised concerning how content generated by bots or AI alters the political make-up of social media feeds. Many studies have shown that social media bots amplify political content [150], [152], [158], [159]. However, once again it is not clear whether this altered exposure drives polarization. Social media platforms have been under significant pressure to remove these bots, and other content believed to be problematic, including mis-/disinformation or hate speech [18], [113], [114]. However, the ability to identify this content requires appropriate detection tools, and the efficacy of these tools is language, and therefore country, dependent. For example, recent research shows that Google's toxicity detection tool "Perspective" is far more likely to label content in German as toxic, than content in other languages [160]. These challenges may be further exacerbated as generative AI improves and it becomes harder to distinguish between machine- and user-generated content [161].

A key factor determining how social media companies moderate their platforms is the country-specific regulatory environments in which they operate [162], [163]. Currently, state regulators can issue takedown requests to remove content they deem to be problematic at a national level (see for example, [help.twitter.com/en/rules-and-policies/post-withheld-by-country](help.twitter.com/en/rules-and-policies/post-withheld-by-country)). However, in most cases regulators cannot remove this content without using the – typically U.S.-based – social media companies as an intermediary. Furthermore, there have been numerous cases where takedown requests have been rejected, e.g., in India [164]. This state of affairs puts an outsized emphasis on U.S. regulators, U.S. political interests, and a U.S. philosophy on platform regulation.

Regulators may make moderation decisions in the interest of preventing online harms, but they could equally be used to manipulate online discourses and support the political regime in power [33], [165], [166]. Either way, these choices are fundamentally political and rest on value judgments which differ between countries. This has led to a growing desire for digital sovereignty [167], which has benefited platforms outside the Western regulatory sphere.

The outsized emphasis on U.S. interests has also impacted the quality of polarization research. Notably, Meta's U.S. election studies – the gold standard for acquiring causal evidence into the online drivers polarization [6], [11], [12] – are part of a recent academic-industry collaboration. But such collaborations are currently dependent on the goodwill of, and the parameters set by, social media companies. These companies have incentives to prioritize U.S. research, which they pursue by granting researchers some "independence by permission" [13]. Consequently, the Meta studies only provide evidence for the U.S., and only involve U.S.-based researchers. Other scientists may use more creative methods to causally estimate the role of social media in driving polarization (see for example refs. [155], [168]), but without the ability to alter a social platform's algorithms, and with no passive consumption data, the questions which such studies can answer are limited.

These systemic biases are a barrier to diversifying polarization research leading some to call for social media research to be democratized [8], [9]. The European Union's new Digital Services Act (DSA) has potential to transform access to online platform data, enabling a new phase of in-depth research. However, early evidence suggests poor compliance with the DSA by tech companies [10], and for researchers outside Europe and the U.S. barriers to data access will remain.

40. **"However, the current dominance of U.S. research has put an outsized emphasis on developing interventions for affective polarization between Democrats and Republicans in the U.S.."**

Other approaches which have proven effective at reducing partisan animosity in the U.S. include promoting intergroup contact [169], emphasizing a shared identity between partisans [170], meaningful, non-political conversations across party lines [143], [171], and inoculating against polarizing content [172], [173]. However, existing approaches often suffer from not being both durable (lasting for extended periods) and scalable (easily and effectively distributed)[4], and in some cases interventions counter-intuitively increase polarization, or result in other negative side-effects (see Box 1 in ref. [4]; also refs. [144], [148], [174]).

41. **"To prevent further global disparities in polarization research, we argue that researchers must push back against growing restrictions on data access now. Since 2023, several social media companies (including Meta, X, Reddit) have shuttered academic access to their platforms' data, and new opportunities for data access such as the European Union's Data Services Act remain largely unproven."**

See discussions on the need for data availability, and an initial assessment on the success of the Digital Services Act, in Refs. [8], [9], [10].

# Supplementary References